\documentclass[12pt,amsmath,amssymb]{iopart}

\usepackage{epsfig}
\usepackage{graphicx,psfrag,xspace}
\usepackage{graphicx,color}
 \graphicspath{{./}{./figures/}}

\newcommand{\beq}{\begin{equation}}
\newcommand{\eeq}{\end{equation}}
\newcommand{\be}{\begin{equation}}
\newcommand{\ee}{\end{equation}}
\newcommand{\bea}{\begin{eqnarray}}
\newcommand{\eea}{\end{eqnarray}}

\begin{document}

\title{The convex hull for a random acceleration process in two dimensions}

\author{Alexis Reymbaut, Satya N. Majumdar, Alberto Rosso}
\address{CNRS - Universit\'e Paris-Sud, LPTMS, UMR8626 - B\^at.~100, 91405 
Orsay Cedex, France}

\begin{abstract}
We compute exactly the mean perimeter $\langle L(T)\rangle$ and the mean area $\langle A(T)\rangle$
of the convex hull of a random acceleration process of duration $T$ in two dimensions.
We use an exact mapping that relates, via Cauchy's formulae, the computation of the 
perimeter and the area of the convex hull of an arbitrary two dimensional stochastic 
process $[x(t),y(t)]$ to the computation of the extreme value statistics of the
associated one dimensional component process $x(t)$. The latter can be computed
exactly for the one dimensional random acceleration process even though the
process in non-Markovian. Physically, our results are relevant in describing the average 
shape of a semi-flexible ideal polymer chain in two dimensions.

\end{abstract}

\section{Introduction}

\begin{figure} 
\centerline{\includegraphics[width=10cm]{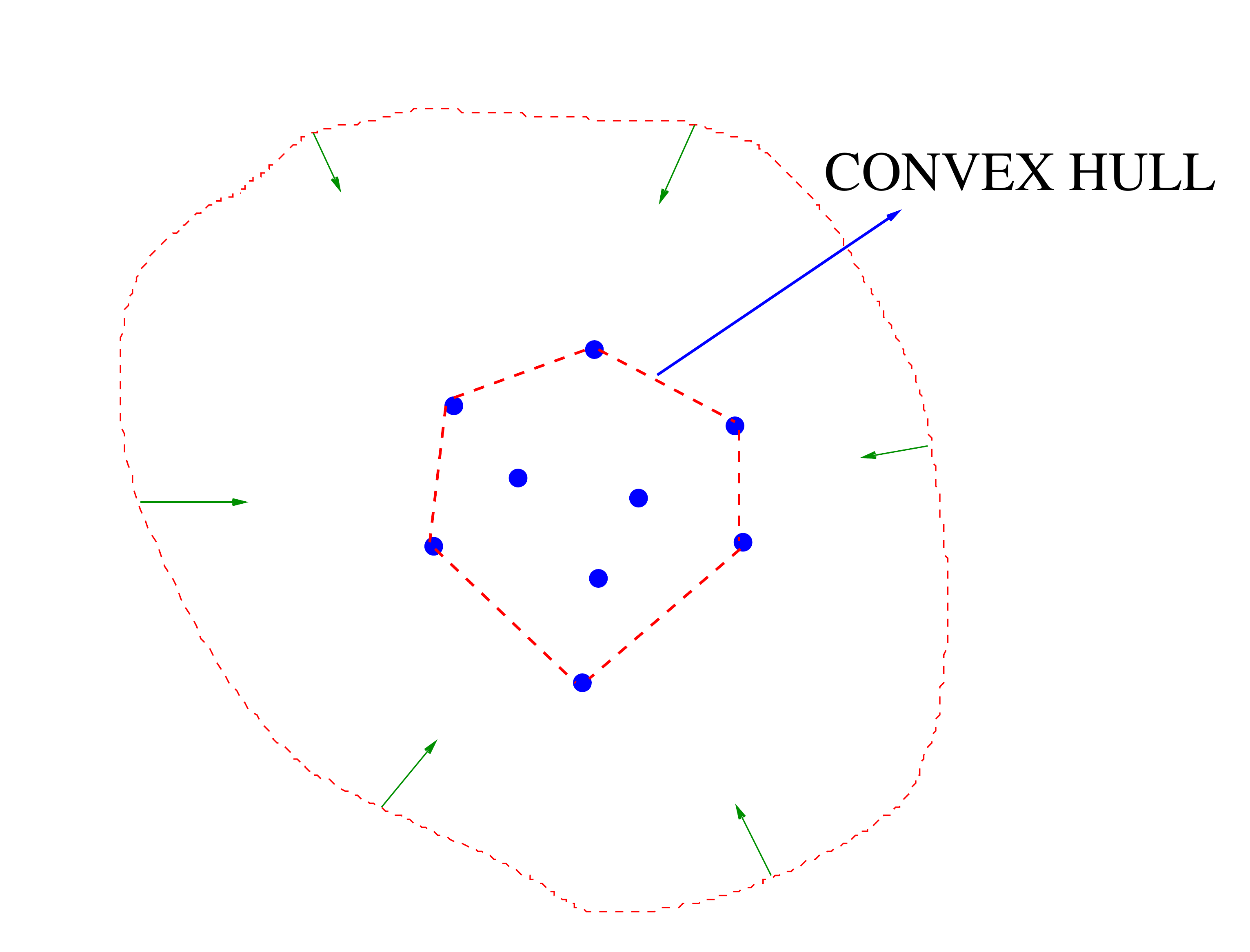}}
\caption{Convex hull of a set of $9$ points obtained by shrinking
an elastic, initially stretched outside the set of points, up to the point 
where it can shrink no more. The resulting final polygon, shown by dotted
red line, is the convex hull associated with this set of points. }
\label{elastic:fig1}
\end{figure}

Consider a set of $N$ points with positions $\{{\vec r_1}, {\vec r_2},\ldots, 
{\vec r_N}\}$ on a two dimensional plane. How does one characterize the
{\em shape} of this set? A standard characterization of the shape
of this set is done by constructing the convex hull associated with this set.
This is simply done in the following way. Let us imagine these points
as nails stuck on a two dimensional board. Let us take a closed elastic or
a rubber band, stretch it so that in includes all the points and
then release it and let it shrink (see Fig. \ref{elastic:fig1}). The
elastic will shrink till it tightly encloses the points and can shrink
no more (see Fig. \ref{elastic:fig1}). At this point, the shape
of the elastic resembles a polygon and this is called the convex
polygon or convex hull associated with the set. It is called convex
because of the property that the line segment joining any two points
on this polygon is fully contained within the polygon.

Properties of convex hulls have been widely studied in mathematics, computer 
science and also in the physics of crystallography in connection with the
so called Wulff construction. Convex hulls are widely used in computer aided 
image processing, in particular for pattern recognition~\cite{Akl}. Finding 
efficient 
algorithms to construct the convex hull of a set of points has led to
numerous studies over the last few 
decades~\cite{Graham,Jarvis,Eddy,Devroye,Kirkpatrick,Wenger}. In addition, 
convex hulls
are used as a standard estimator for the home range for animal movements in
ecology, leading to several important practical 
applications~\cite{Sirakov,Worton,Luca}. For a
recent review on the history and applications of convex hulls, see 
Ref. \cite{extremv}.

Of particular interest are the statistical properties of the convex hull
associated with a set of random points on the plane. Consider, for instance,
the set of $N$ points whose coordinates $\{{\vec r_1}, {\vec r_2},\ldots,
{\vec r_N}\}$ are random variables drawn, in general, from a
joint distribution $P\left({\vec r_1}, {\vec r_2},\ldots,
{\vec r_N}\right)$. For each realization of the points, one 
can construct the associated convex hull and compute observables such as the
perimeter $L$, area $A$ or the number of vertices $V$ of this convex hull.
Clearly these observables change from one realization of points to another
and are themselves random variables. Given the underlying distribution
of the points $P\left({\vec r_1}, {\vec r_2},\ldots,
{\vec r_N}\right)$, a challenging hard problem is to compute the statistics
of the observables associated with the convex hull, e.g., 
the distributions such as $P(L)$, $P(A)$ or $P(V)$. Even computing
the mean perimeter $\langle L\rangle$ , mean area $\langle A \rangle $ and the 
mean number of vertices $\langle V \rangle $ on the
convex hull is, in general, a hard problem. 
This is not just  a mathematically challenging problem, but has 
several important practical applications, as reviewed recently
in Ref. \cite{extremv}. 
While several results 
are known (see \cite{extremv} for a review) in the case when the points are 
distributed independently
and identically, i.e., when the joint distribution factorises, $P\left({\vec 
r_1}, {\vec r_2},\ldots,
{\vec r_N}\right)= \prod_{i=1}^N p(\vec r_i)$, very few results are
known when the vertices are {\em correlated}. 

A classic example when the points are correlated is the case when 
${\vec r_i}$ represents the position of a two-dimensional random walk
at time step $i$. Each component of the position of the random walker evolves 
via the Markov evolution rule
\begin{eqnarray}
x_i &= &x_{i-1}+ \eta_x(i) \label{rwx} \\ 
y_i &= &y_{i-1}+\eta_y(i) \label{rwy}
\end{eqnarray}
starting from $x_0=y_0=0$ where the jump lengths $\eta_x(i)$ and $\eta_y(i)$
in the $x$ and $y$ directions are random variables, independent from each other 
and are
also independent from one step to another and each is a zero mean Gaussian
variable with a finite variance $\sigma^2$. The walk evolves upto $N$
steps and one can ask, for instance, what is the mean perimeter and the mean 
area of the convex hull associated with this random walk trajectory?
These results are known in the continuous-time limit
when $N\to \infty$ and $\sigma^2\to 0$ keeping
the product $N\sigma^2= 2DT$ fixed,  
where $T$ is the total duration of the walk and $D$ is the diffusion constant.
In the continuous-time limit, Eqs. 
(\ref{rwx})-(\ref{rwy}) reduce to the Brownian motion in a plane
\begin{eqnarray}
\frac{dx}{dt}&=&\eta_x(t) \label{bmx} \\
\frac{dy}{dt}&=&\eta_y(t) \label{bmy}
\end{eqnarray}  
starting at the origin $x(0)=y(0)=0$, and
where $\eta_x(t)$ and $\eta_y(t)$ are Gaussian white noises with zero mean
and the two-time correlators, $\langle \eta_x(t)\eta_x(t')\rangle = 2D 
\delta(t-t')$, $\langle \eta_y(t)\eta_y(t')\rangle =
2D \delta(t-t')$ and $\langle \eta_x(t)\eta_y(t')\rangle =0$.
In this
continuous-time limit, the mean perimeter of the
convex hull associated to this Brownian motion of total duration $T$ was first 
computed by Tak\'acs, $\langle L(T)\rangle =\sqrt{16\pi\, D\,T}$~\cite{Takacs}.
Note that while the $\sqrt{D\,T}$ dependence of the perimeter is expected
due to the diffusive nature of the path, the computation of the
prefactor $\sqrt{16\pi }$ is highly nontrivial. Later, the mean perimeter
of the convex hull associated with a two dimensional Brownian bridge
(where the walker is constrained to come back to the origin after time
$T$) was also computed exactly by Goldman, $\langle L(T)\rangle_{\rm 
bridge}=\sqrt{\pi^3 D T}$~\cite{Goldman}. The mean area of the convex hull
for a free Brownian motion of duration $T$ , $\langle A(T)\rangle = \pi\,D\,T$,  
was first computed by El Bachir and Letac~\cite{ElBachir,Letac}, while the 
corresponding 
result for the bridge, $\langle A(T)\rangle_{\rm bridge}= (2\pi/3)\,D\,T$ was
computed only very recently~\cite{RMC}. Once again, while the linear $D\,T$
dependence is expected (as the area $A\sim L^2$), the prefactors
were nontrivial to compute. Moreover, the mean perimeter $\langle 
L_n(T)\rangle=\alpha_n \sqrt{D\,T}$
and the mean area $\langle A_n(T)\rangle= \beta_n (D\,T) $ of the convex hull 
of $n$ 
independent two-dimensional Brownian motions each of duration $T$ was
exactly computed recently for all $n$~\cite{RMC,extremv} both for free
Brownian paths and Brownian bridges and the prefactors $\alpha_n$
and $\beta_n$ were found to have nontrivial $n$ dependence.

While the positions of a two-dimensional Brownian motion are correlated, the
time evolution of the process is still {\em Markovian}. An important 
question is if one
can compute the statistics of the convex hull associated to a 
two-dimensional {\em 
non-Markovian} stochastic process. This would then be a nontrvial 
generalisation.
The purpose of this paper is to present exact analytical results for
the mean perimeter and the mean area of the convex hull associated with
the trajectory of a two-dimensional {\em non-Markovian} stochastic process, 
namely, the two dimensional random acceleration process. In this process,
the position $(x,y)$ of a particle in the plane evolves in continuous-time 
$t$ via 
\bea 
\frac{d^2 x}{ d t^2}&= &\eta_x(t) \label{rax} \\
\frac{d^2 y}{ d t^2}& = & \eta_y(t) \label{ray}
\eea
where $\eta_x(t)$ and $\eta_y(t)$ are zero mean  
Gaussian white noises as in Eqs. (\ref{bmx})-(\ref{bmy}), with
two-point correlators, $\langle \eta_x(t)\eta_x(t')\rangle = 2D
\delta(t-t')$, $\langle \eta_y(t)\eta_y(t')\rangle =
2D \delta(t-t')$ and $\langle \eta_x(t)\eta_y(t')\rangle =0$.
Henceforth, for simplicity, we will set $D=1$ without any loss
of generality. Note that due to the presence of the second derivative in 
Eqs. (\ref{rax})-(\ref{ray}), the evolution is non-Markovian. If however, one defines the
velocities, $v_x= dx/dt$ and $v_y=dy/dt$, then in the position-velocity phase space, the
evolution equation involves only first derivative in time and hence the joint $(\vec r,\vec v)$
process becomes Markovian,
\bea
\frac{dx}{dt}&= & v_x(t); \quad\quad  \frac{dv_x}{dt}= \eta_x(t) \label{raxvx} \\
\frac{dy}{dt}& = & v_y(t); \quad\quad  \frac{dv_y}{dt}= \eta_y(t). \label{rayvy}
\eea

\begin{figure} 
\centerline{\includegraphics[width=10cm]{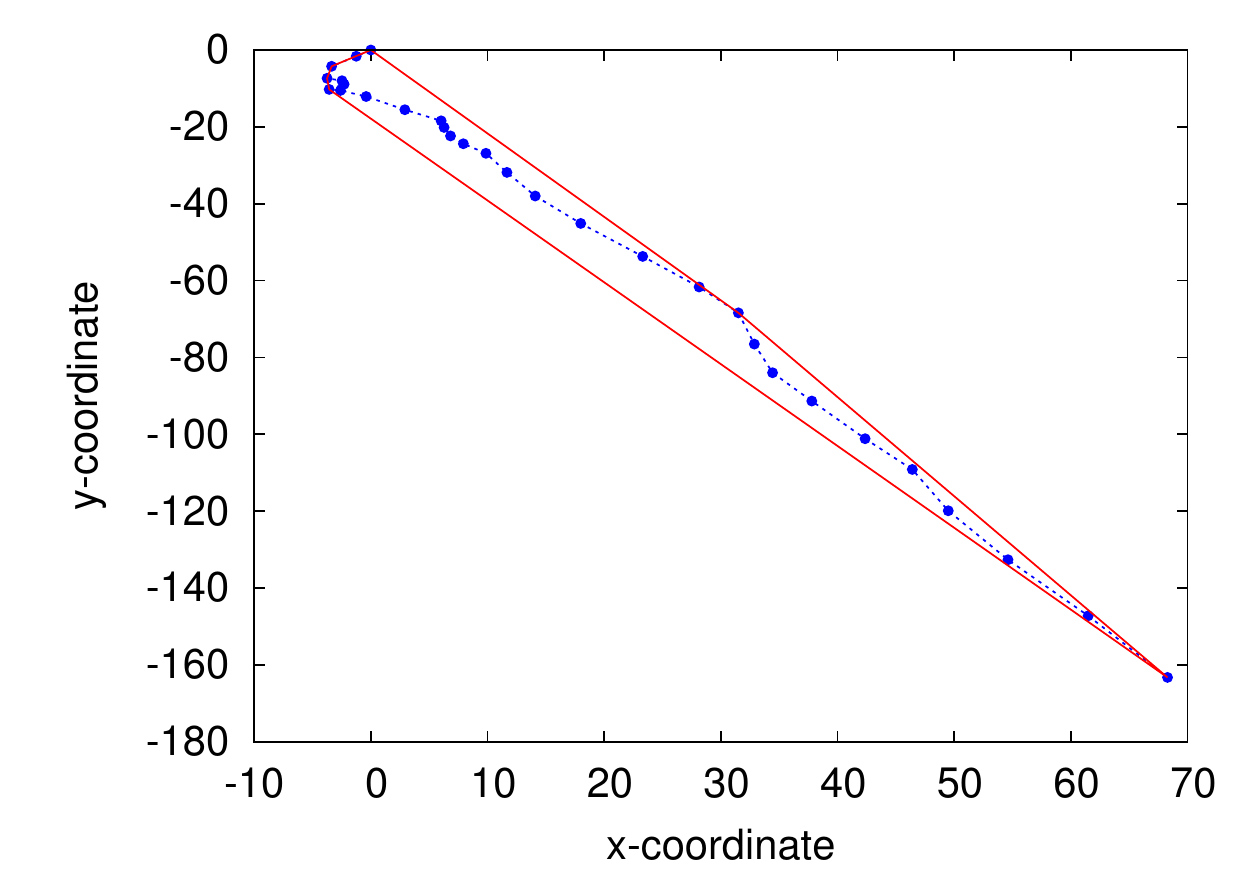}}
\caption{Convex hull of $30$ points generated using the  
discrete random acceleration process defined in Eqs. (\ref{discrete1})-(\ref{discrete2}). 
The convex hull is constructed using the Graham span algorithm~\cite{Graham}.}
\label{typical.fig}
\end{figure}

At time instant $t$ 
the particle is identified by its position $\vec{r}(t)=(x(t),y(t))$
and its velocity 
$\vec{v}(t)=\left(v_x(t)=dx/dt, v_y(t)=dy/dt\right)$.
We assume that the particle starts at the origin $(x=0,y=0)$ at time $0$
with initial zero velocities: $(v_x(0)=0, v_y(0)=0)$. We evolve the
process up to time $T$ and then construct the convex hull associated
with the trajectory in the $(x-y)$ plane and compute the statistics of 
observables
such as the perimeter $L(T)$, the area $A(T)$ and the number of vertices 
$V(T)$.
A particular realization of this process and the associated convex hull is 
shown in Fig. (\ref{typical.fig}).  

Our main results are summarized as follows. We show that the mean perimeter
and the mean area of the convex hull associated to a two-dimensional
random acceleration process of duration $T$, starting from the 
initial zero velocity conditions $v_x(0)=v_y(0)=0$,
are given by the following exact expressions
\bea 
\langle L(T) \rangle &=&  \sqrt{\frac{3 \pi}{2}}\,  T^{3/2} =(2.1708\ldots )\, 
T^{3/2} \label{ramper1}
\\
 \langle A(T) \rangle &=& \frac{ 5 \pi}{192}\,   \sqrt{\frac{3}{2}}\,  T^3
= (0.100199\ldots)\, T^3. 
\label{ramarea1}
\eea
As in case of Brownian motion, while the $T$ dependence of the mean perimeter
and the mean area easily follow from dimensional grounds, computation of the 
prefactors turns
out to be rather nontrivial and they are the main results of this paper.
We could not compute the mean number of vertices analytically, for which
we present only numerical results.

The random acceleration process is perhaps the simplest non-Markovian 
process and hence, no wonder, that it has been intensively studied
both in the Physics and the Mathematics literature. Even though it is simple,
the first-passage properties of this process are highly nontrivial
even in one dimension, though several exact results are known
~\cite{McKean,MGoldman,MW,Sinai,Burkhardt,Lachal,MB1,Burkhardt2,GL,MB2,Gyorgyi,SM1,
Burkhardt3,Burkhardt4,Burkhardt5,time}. There are several applications of this 
process, notably
in the continuum description of semiflexible polymer chain~\cite{Burkhardt},
of fluctuating linear interfaces with dynamic exponent $z=4$~\cite{MB2}
and also in the description of the statistical properties of Burgers equation 
with Brownian initial velocity~\cite{PV}. Thus, apart from being 
mathematically interesting, our results in this paper concerning 
the statistical
properties of the convex hull of a two-dimensional random acceleration process
are physically relevant in describing the average `shape' of 
a ideal semi-flexible polymer chain in two dimensions (without the
excluded volume interaction and with one end fixed at the origin while
the other end is free).  

In deriving our exact results, we map the problem of computing the perimeter
and the area of the two dimensional random acceleration process to the
problem of computing the statistics of extremes of a one dimensional
random acceleration process, following the general method introduced 
recently in \cite{RMC,extremv}. We remark that a similar mapping
between the convex hull of a set of independent points drawn from a
unit disc and an effective one dimensional random acceleration process
was also noticed recently in the context of the so called Sylvester's 
question~\cite{Hilhorst}. The Sylvester's question asks: given a
set of $N$ independent points (drawn from a unit disc), what is the
probabablity that all $N$ points lie {\em on} the convex hull?
In other words, in our notation the probability $P(V=N)$ where $V$
is the number of vertices {\em on} the convex hull. Based on the mapping
to the one dimensional random acceleration process, the authors
of Ref. \cite{Hilhorst} derived exact asymptotic results for $P(V=N)$
for large $N$.

The rest of the paper is organized as follows. In Section 2, we 
show how one can use Cauchy's formulae to map the problem of computing the perimeter and the
area of the convex hull of a generic two dimensional stochastic process
to the problem of computing the moments
of the maximum and the time at which the maximum occurs for the
associated one dimensional component stochastic process. We then focus
on the random acceleration process and show how to compute explicitly
the mean perimeter and the mean area of its convex hull using this mapping.
In Section 3, we present the
results of numerical simulations. We conclude in Section 4 with
a summary and open problems. The detailed 
calculation of the moments of the maximum of a one dimensional random accelertion process,
required for our main results, is relegated to the Appendix A.

\section{Mapping to the extreme value statistics of a one dimensional process
using Cauchy's formulae}

In Refs. \cite{RMC,extremv} it was shown that the problem of computing
the perimeter and the area of the convex hull of any two dimensional
stochastic process $[x(t),y(t)]$ can be mapped to computing the statistics of 
the
value of the maximum and the time of occurrence of the maximum of
the one dimensional component process $x(t)$. This was achieved in 
\cite{RMC,extremv} by exploiting the formulae for the perimeter
and the area of any arbitrary closed, convex curve in two dimensions,
derived originally by Cauchy~\cite{Cauchy}.
We start this section by briefly reviewing
this useful mapping and then use this mapping to compute explicitly the mean
perimeter and the mean area of the convex hull of the two dimensional
random acceleration process defined in Eqs. (\ref{rax})-(\ref{ray}).   

\begin{figure}
\centerline{\includegraphics[width=10cm]{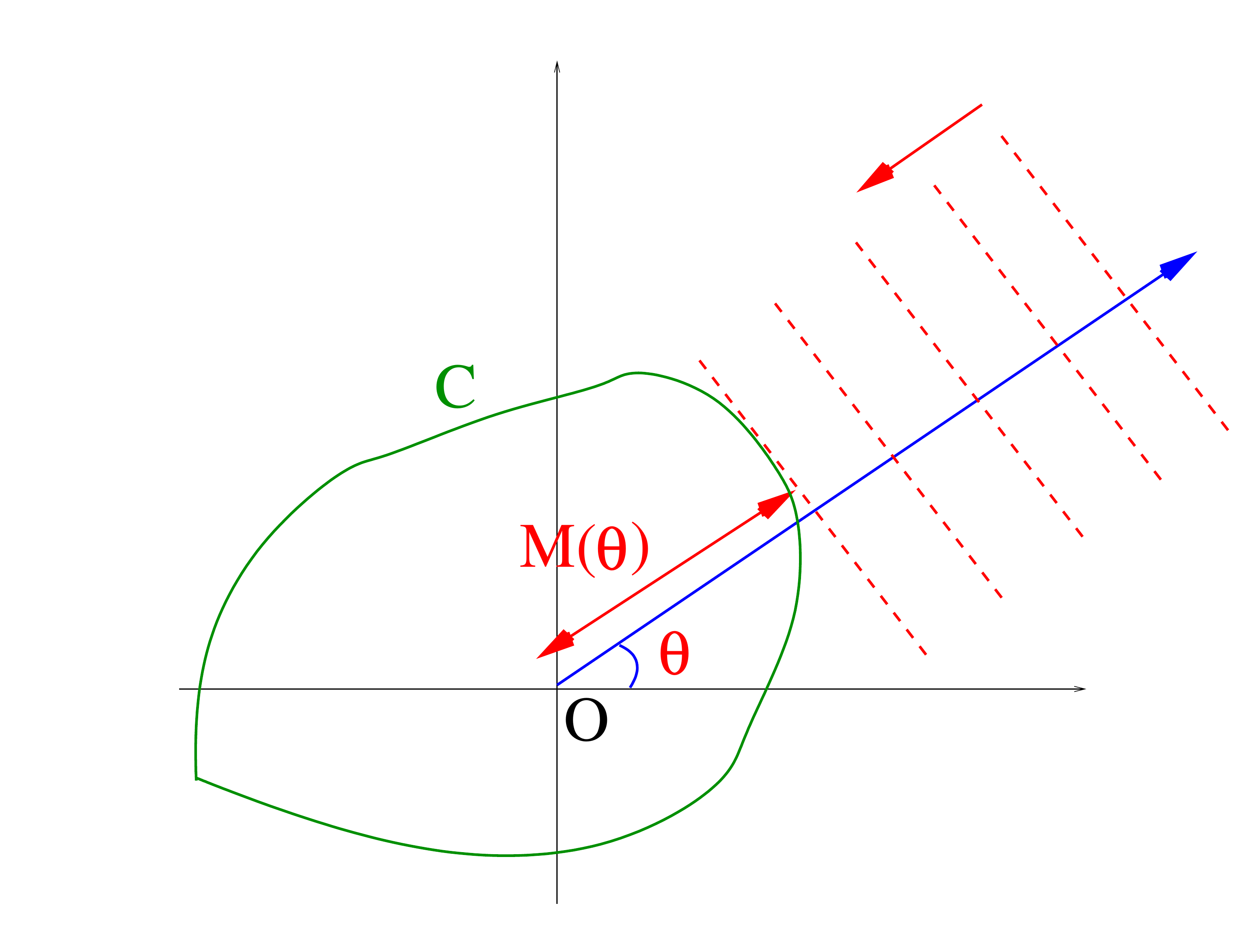}}
\caption{Support function $M(\theta)$ in the direction $\theta$ for
a closed convex curve $C$ on a plane.}
\label{support.fig}
\end{figure}

Consider any closed convex curve $C$ on a plane (see Fig. \ref{support.fig}).
We choose the coordinate system such that the origin is inside the curve $C$.
A key quantity, introduced by Cauchy~\cite{Cauchy},
is the so called support function. Consider any direction $\theta$. For fixed 
$\theta$, consider a stick perpendicular to this direction and imagine
bringing the stick from infinity and stop when it first touches
the curve $C$ (see Fig. \ref{support.fig}). At this point, the distance
$M(\theta)$ of the stick from the origin is called the support function
in the direction $\theta$. It is clearly a very intuitive quantity
as it measures how close can one get to the curve $C$ in the direction
$\theta$ coming from infinity.
Mathematically speaking, consider all points on the curve $C$ with
coordinates $[X(s),Y(s)]$ (where $s$ is the distance measured along the 
curve that just parametrizes the curve), consider their projections
in the direction $\theta$ and then compute
the maximum projection, i.e., 
\begin{equation}
M(\theta)= \max_{s\in C}\left[X(s)\cos(\theta)+ Y(s)\sin(\theta)\right].
\label{support1}
\end{equation}
Once this support function $M(\theta)$ is known, then the perimeter $L$
of $C$ and the area $A$ enclosed by $C$ are given by Cauchy's 
formulae~\cite{Cauchy}
\begin{eqnarray}
L & =& \int_0^{2\pi} M(\theta)\, d\theta \label{Cperi} \\
A &= & \frac{1}{2} \int_0^{2\pi}\left[M^2(\theta)- (M'(\theta))^2\right]\, 
d\theta
\label{Carea}
\end{eqnarray}
where $M'(\theta)= dM/d\theta$. For example, for a circle of radius $R=r$, 
$M(\theta)=r$ and one recovers the standard formulae: $L=2\pi r$ and $A=\pi 
r^2$. The formulae in Eqs. (\ref{Cperi})-(\ref{Carea}) are generalisations
of the circle formulae, valid for arbitrary closed convex curve $C$. A simple
proof of the formulae can be found in the Appendix A of Ref. \cite{extremv}.

Now, consider any two dimensional stochastic process $[x(t),y(t)]$ up to time 
$T$, starting from $[x(0)=0, y(0)=0]$. For every realization, i.e., for every 
trajectory of the process, let us construct the corresponding convex hull $C$, 
which clearly varies from one trajectory to another. The idea then is to apply
the Cauchy's formulae to each of these closed convex curves $C$ and
then average over all realizations to compute the mean perimeter and
the mean area. To apply Cauchy's formulae to a particular convex hull $C$, we 
need to first compute its associated support function $M(\theta)$ defined
in Eq. (\ref{support1}). It looks rather complicated to first compute
the locus $[X(s),Y(s)]$ of the convex hull and then maximize over $s$. However, the
crucial point to realize~\cite{RMC,extremv} is that $M(\theta)$ represents the 
maximum projection
of the convex hull $C$ in the direction $\theta$, which is also precisely
the maximum projection of the trajectory $[x(t),y(t)]$ itself in the
direction $\theta$, i.e.,
\begin{eqnarray}
M(\theta) &= &\max_{s\in C}\left[X(s)\cos(\theta)+ Y(s)\sin(\theta)\right] 
\nonumber \\
& =& \max_{0\le t\le T}\left[x(t) \cos(\theta)+y(t)\sin(\theta)\right].
\label{support2}
\end{eqnarray}
So, in principle, if we have the list of cordinates visited by the
trajectory up to time $T$, we can compute the support function $M(\theta)$
by maximizing over time $t$. Let us denote $z_\theta(t)= x(t)\cos (\theta)
+ y(t)\sin (\theta)$. For a fixed $\theta$, $z_{\theta}(t)$ is
just a one dimensional process parametrized by $\theta$.
Thus the support function $M(\theta)$ is just the maximum
of this one dimensional process $z_{\theta}(t)$ over the 
time interval $t\in [0,T]$.   
Furthermore, assuming that the maximum of 
$z_{\theta}(t)$ (with $0\le t\le T$) 
occurs at time $t_m$, we have
\begin{equation}
M(\theta)= \max_{0\le t\le T}\left[z_{\theta}(t)\right]= z_{\theta}(t_m)= 
x(t_m)\cos (\theta) + y(t_m) \sin(\theta).
\label{max1}
\end{equation}
where $t_m$ implicitly depends on $\theta$. The derivative 
$M'(\theta)$ needed in the area formula in Eq. (\ref{Carea}) can then be computed by 
deriving the right hand side (r.h.s) of Eq. (\ref{max1}) with respect to $\theta$. This gives
\begin{equation}
M'(\theta)=- x(t_m) \sin(\theta)+ y(t_m)\cos (\theta) + 
\frac{dt_m}{d\theta}\,\frac{dz_{\theta}(t)}{dt}\Big|_{t=t_m}. 
\label{max2}
\end{equation}
However, since $z_{\theta}(t)$ is maximum at $t=t_m$, by definition, 
$dz_{\theta}(t)/dt|_{t=t_m}=0$ and hence the
last term on the r.h.s vanishes and one simply gets
\begin{equation}
M'(\theta)= - x(t_m) \sin(\theta)+ y(t_m)\cos (\theta). 
\label{dmax}
\end{equation}

\begin{figure} 
\centerline{\includegraphics[width=8cm]{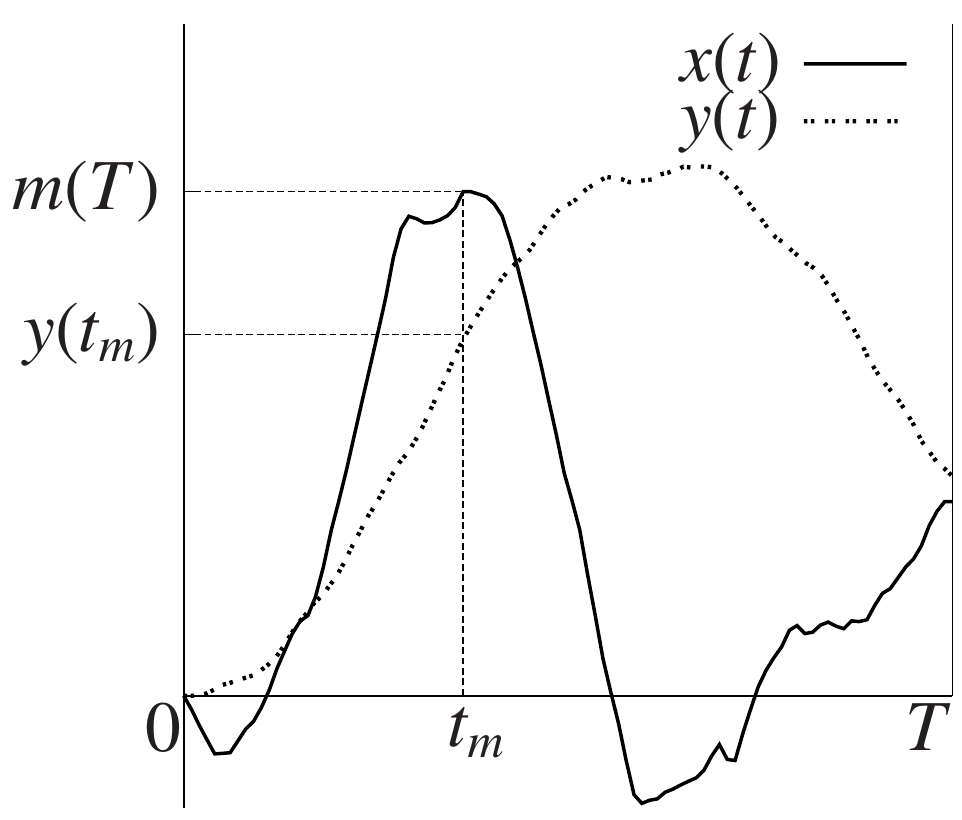} }
\caption{Schematic picture of $m(T)$, $t_m$ and $y(t_m)$.  }
\label{sketch}
\end{figure}

The equations (\ref{max1}) and (\ref{dmax}) are the essential ingredients
behind the mapping of the original two dimensional convex hull problem
to the statistics of the maximum of an effective one dimensional
process. Now, an additional simplicity can arise if the 
two dimensional stochastic process is rotationally invariant, as in
the case of Brownian motion and also for the random acceleration process.
In this case, any average over the realizations is independent
of the angle $\theta$. Thus, the integral $I=\int_0^{2\pi} \langle F[M(\theta)] 
\rangle\, d\theta $ just reduces, thanks to the isotropy, to $I=2\pi \langle 
F[M(0)]\rangle$, for any function $F[M]$.
Setting $\theta=0$ in Eqs. (\ref{max1}) and (\ref{dmax}) we note that
\begin{eqnarray}
M(0) &= & x(t_m)= m (T) \label{m0}\\
M'(0) &= & y(t_m) \label{m'0}
\end{eqnarray}
where $m(T)$ denotes the maximum of the one dimensional component process 
$x(t)$
over the interval $0\le t\le T$ and $t_m$ denotes the time at which
$x(t)$ achieves its maximum. On the other hand $M'(0)$ is the value
of the other independent one dimensional process $y(t)$ at the time
$t=t_m$ when the first process $x(t)$ attains its maximum (see Fig. 
(\ref{sketch})).

Thus, the mean perimeter and the mean area,
taking average over Cauchy's formulae, using isotropy, and Eqs. (\ref{m0})
and (\ref{m'0}), are simply given by
\begin{eqnarray}
\langle L(T)\rangle &=& 2\pi\, 
\langle M(0)\rangle = 2\pi \langle m(T)\rangle \label{avper} \\
\langle A(T)\rangle &=& \pi\left[\langle 
M^2(0)\rangle - 
\langle (M'(0))^2\rangle \right]= \pi \left[\langle m^2(T)\rangle-\langle 
y^2(t_m)\rangle\right] \label{avarea}
\end{eqnarray} 

So far, the results in Eqs. (\ref{avper}) and (\ref{avarea}) are quite general
and hold for any arbitrary two dimensional rotationally invariant stochastic 
process $[x(t),y(t)]$.
To compute the mean perimeter and the mean area, we then need to compute
three quantities for the associated one dimensional component process $x(t)$, 
namely the first two moments of the maximum $m(T)$ of
the one dimensional process $x(t)$ over the interval $0\le t\le T$: (i) $\langle 
m(T)\rangle$ 
(ii) $\langle m^2(T)\rangle $ and also (iii) $\langle y^2(t_m)\rangle$ where
$t_m$ is the time at which the process $x(t)$ attains its maximum value.
For the Brownian motion, these three quantities can be easily 
calculated~\cite{RMC,extremv}. Here we show that for the random acceleration 
process as well, one can compute these three quantities explicitly.

For a one dimensional random acceleration process
$x(t)$, evolving via $d^2x/dt^2=\eta_x(t)$, over the interval $0\le t\le T$,
the Laplace transform of the distribution of the maximum was obtained recently by 
Burkhardt~\cite{Burkhardt5}. This Laplace transform has a rather complicated
form (expressed as an integral over Airy function) (see Eq. (A.4) in Appendix A).
Computing explicitly the moments of $m(T)$ from
this Laplace transform turns out to be rather nontrivial. However, fortunately this can be 
done as we show in Appendix A. In particular, for the first two moments, we
get
\bea
\langle m(T) \rangle = \sqrt{\frac{3}{8 \pi}}\, T^{3/2} \label{mom1}\\
\langle m^2(T) \rangle =\left(  \frac{2}{3} -
\frac{103}{384} \sqrt{\frac{3}{2}}   \right) \, T^3. \label{mom2}
\eea
Finally, it rests to compute $\langle y^2(t_m)\rangle$. This can be done
in two steps. First, we note from the evolution equation, $d^2y/dt^2=\eta_y(t)$
[where $\langle \eta_y(t)\rangle=0$ and $\langle \eta_y(t)\eta_y(t')\rangle 
=2\delta(t-t')$], that at any fixed time $t$, $\langle y^2(t)\rangle =(2/3)\, 
t^3$
for all $t$. This is easily obtained by integrating the evolution equation
and computing $\langle y^2(t)\rangle$ using the delta correlation of
the noise. Secondly, averaging over the distribution $p(t_m|T)$ of the
time of the maximum $t_m$ of $x(t)$ one gets
\begin{equation}
\langle y^2(t_m)\rangle = \frac{2}{3} \langle t_m^3\rangle=
\frac{2}{3}\, \int_0^T dt_m\, t_m^3\, p(t_m|T). \label{tm3}
\end{equation}  
So, we need to know the distribution $p(t_m|T)$ of the time $t_m$ at which
a random acceleration process achieves its maximum over the interval
$t\in [0,T]$. 
Fortunately, this is
exactly known from our recent calculation~\cite{time}
\begin{equation}
p(t_m|T)=C\,\delta(T-t_m)+\frac{(1-C)}{\pi \sqrt{2}} t_m^{-3/4}\,(T-t_m)^{-1/4},
\label{result2}
\end{equation}
where the constant $C$ has the exact value
\begin{equation}
C=1-\sqrt{\frac{3}{8}}=0.387628\ldots
\label{resultC}
\end{equation}
Using this result in Eq. (\ref{tm3}) we finally get
\be
\langle y^2(t_m)  \rangle= \frac{2}{3}
\left(C+ \frac{15}{128} (1-C)\right) T^3=
\left(\frac{2}{3}-\frac{113}{384}\sqrt{\frac{3}{2}}\right) T^3.
\label{y2tm}
\ee

Substituting the results from Eqs. (\ref{mom1}), (\ref{mom2}) and (\ref{y2tm})
in Eqs. (\ref{avper}) and (\ref{avarea}) immediately gives our main results
\bea 
\langle L(T) \rangle &=&  \sqrt{\frac{3 \pi}{2}}\,  T^{3/2} =(2.1708\ldots )\,
T^{3/2} \label{ramper2}
\\
 \langle A(T) \rangle &=& \frac{ 5 \pi}{192}\,   \sqrt{\frac{3}{2}}\,  T^3
= (0.100199\ldots)\, T^3.
\label{ramarea2}
\eea

\section{Numerical simulations} \label{numerics}

To verify our main theoretical predictions, we have also performed simulations of a 
discrete version of the random acceleration process in two dimensions. A particle at discrete time, 
$n=0,1,\ldots,N$, is identified by its position $(x(n),\, y(n))$ and its velocity 
$(v_x(n),\, v_y(n))$.  At time $n+1$ the velocity is updated by the discrete-time version
of the Markov evolution rules in Eqs. (\ref{raxvx}) and (\ref{rayvy}):
\bea
   \label{discrete1}
    v_x(n) & = &  v_x(n-1) +  \eta_x(n) \nonumber \\
    v_y(n) & = &   v_y(n-1) +  \eta_y(n) 
   \eea
where $\eta_x(n), \eta_y(n)$ are uncorrelated Gaussian normal variables each with
zero mean and variance $\sigma^2=2$. The position 
$(x(n),\, y(n))$ is then updated:

   \bea
   \label{discrete2}
    x(n) & = & x(n-1) + v_x(n)  \nonumber \\
    y(n) & = & y(n-1)+ v_y(n) 
   \eea
The process is started from $x(0)=y(0)=v_x(0)=v_y(0)=0$ and iterated for $N-1$ 
steps and then the convex hull is constructed using the Graham scan algorithm 
\cite{Graham}. An example is shown in Fig. (\ref{typical.fig}).  The output of each 
realization is the ordered set of the $V$ vertices sorted clockwise from 
$(x_1,y_1)$ to $(x_V,y_V)$. The perimeter and the area of the convex hull can then be 
computed using the formulae:
\bea
L= \sum_{i=1}^V \sqrt{(x_{i+1}-x_i)^2+(y_{i+1}-y_i)^2}  \\
A= -\frac{1}{2} \sum_{i=1}^V \left(x_i y_{i+1} - x_{i+1} y_i \right) \label{surveyor}
\eea 
where we assume $x_{V+1}=x_1$ and  $y_{V+1}=y_1$.  
It is important to note that the minus symbol in Eq. (\ref{surveyor}) is due to 
the clockwise sorting of the points: indeed, the area formula (commonly called the 
surveyor's formula) gives the algebrical value of the area of a polygon; it gives 
the real area if the points are sorted counterclockwise, which is not the case 
here.

\begin{figure} 
\centerline{\includegraphics[width=12cm]{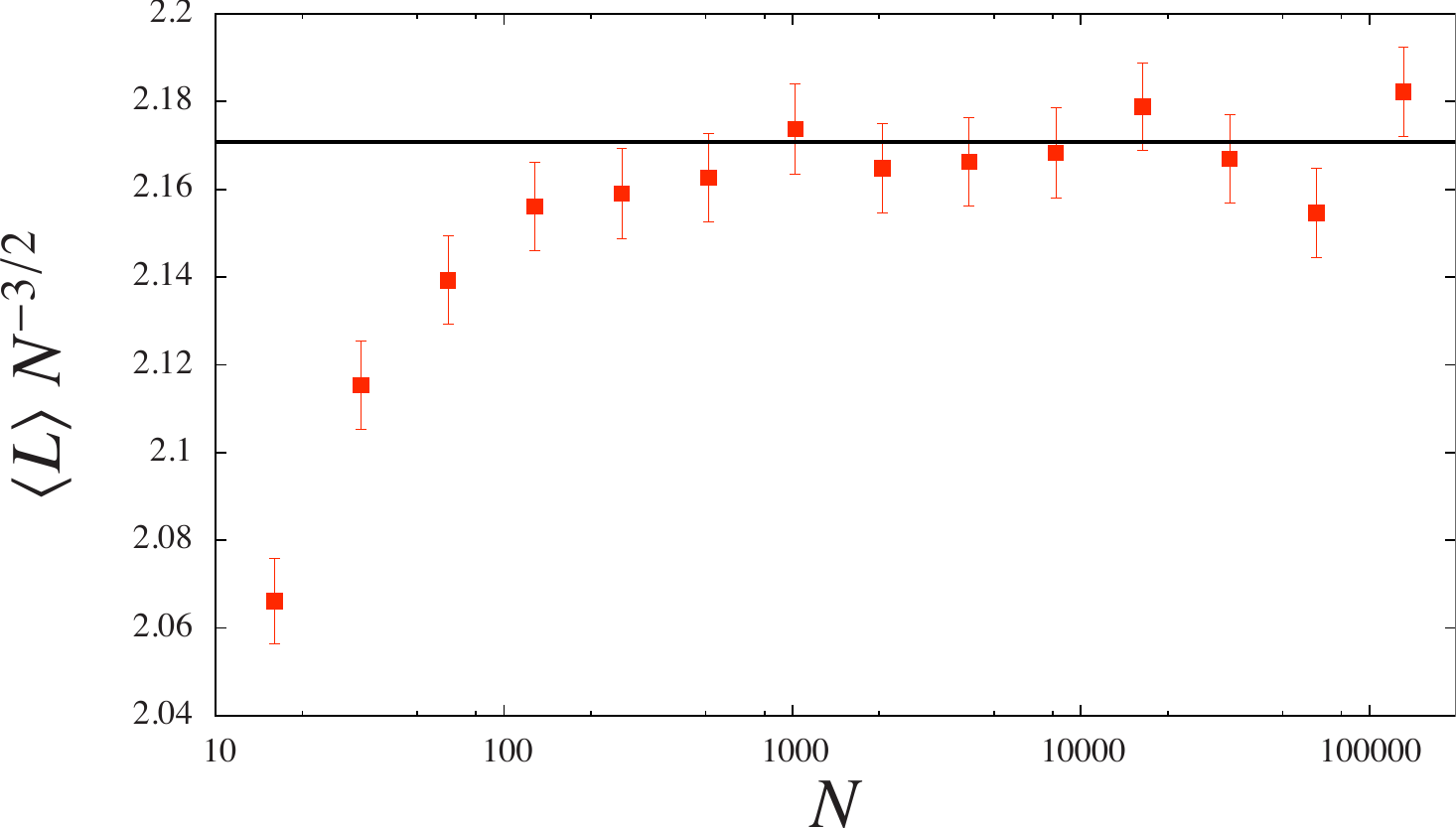} }
\caption{Behavior of $\langle L\rangle$. Symbols: numerical data for the model 
defined by Eqs. (\ref{discrete1}) and (\ref{discrete2}). Solid line: continuum 
model prediction, $\sqrt{3 \pi /2}\sim 2.1708038 \ldots$.  }
\label{figp}
\end{figure}

\begin{figure} 
\centerline{\includegraphics[width=12cm]{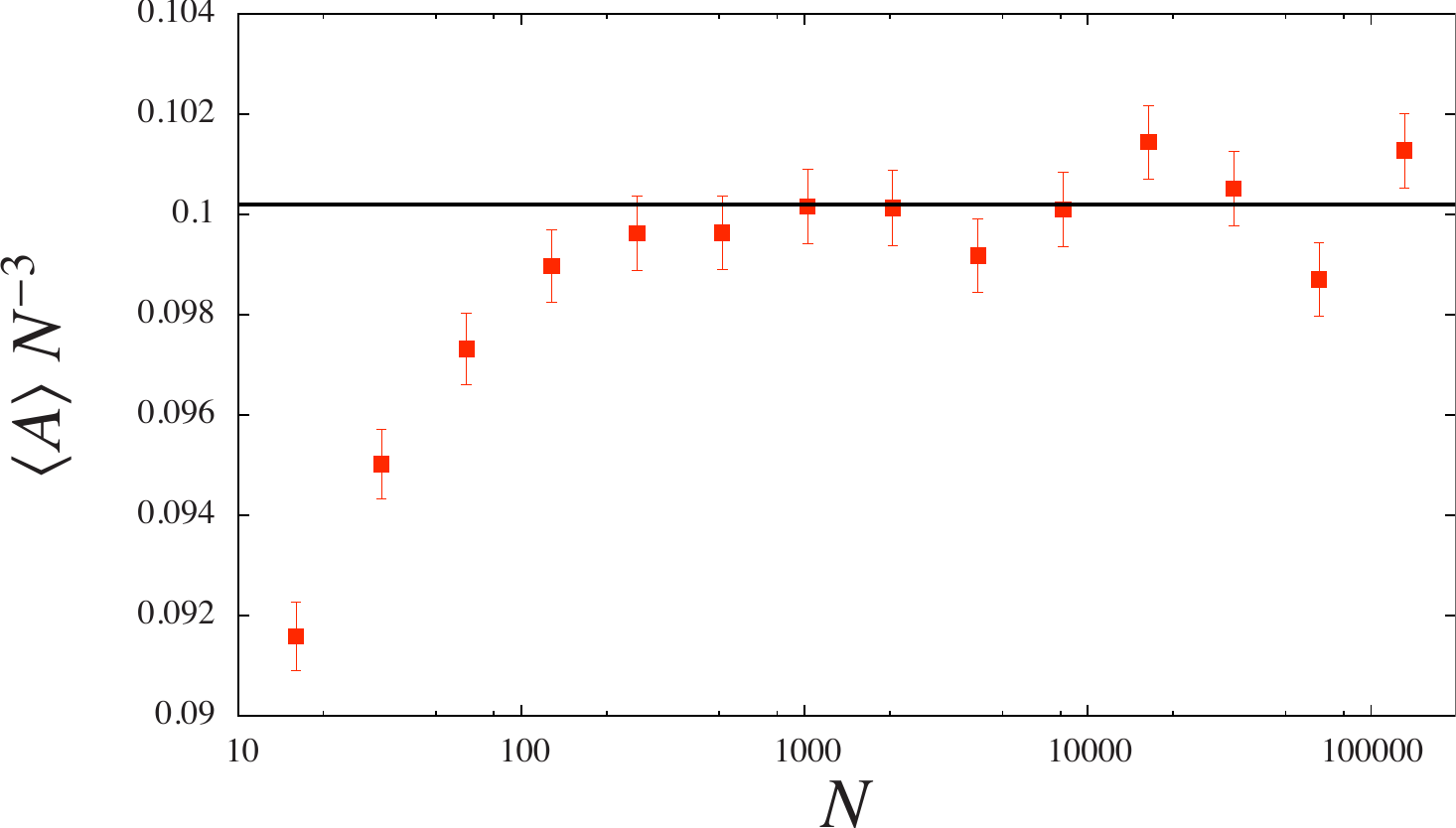}} 
\caption{Behavior of $\langle A\rangle$. Symbols: numerical data for the model 
defined by Eqs. (\ref{discrete1}) and (\ref{discrete2}). Solid line: continuum 
model prediction, $5 \pi \sqrt{3 /2}/192\sim 0.10019921 \ldots$.  } 
\label{figa} 
\end{figure}

\begin{figure}
\centerline{\includegraphics[width=12cm]{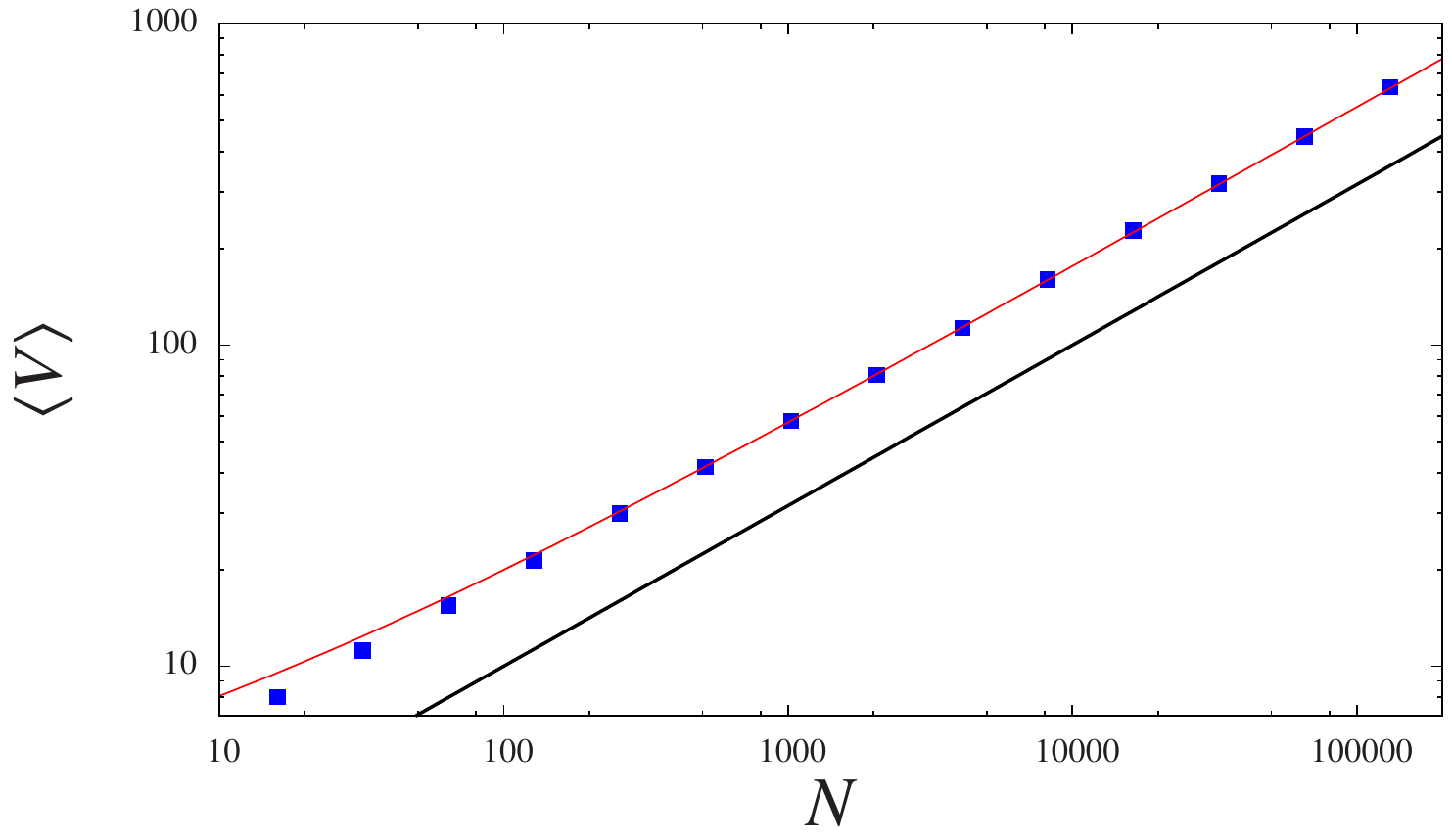}}
\caption{Behavior of $\langle V\rangle$. Symbols: numerical data for the model 
defined by Eqs. (\ref{discrete1}) and (\ref{discrete2}). Solid line: scaling 
behavior $\sim \sqrt{N}$. Dashed line:  best fit.}
\label{figV}
\end{figure}

The perimeter $L$, the area $A$ and the number of vertices $V$ are stochastic 
variables which fluctuate from one realization to another. We computed $\langle 
L\rangle$, $\langle A \rangle$, $\langle V \rangle$ for different values of $N$. 
For each $N$, we average over $10^4$ realizations. For large $N$, the leading 
asymptotic behaviors of the mean perimeter and the mean area should be given by 
the continuous-time limit 
results in Eqs. (\ref{ramper2}) and (\ref{ramarea2})
\bea \label{connection2}
\langle L \rangle &=&  \sqrt{\frac{3 \pi}{2}}\, N^{3/2} +O(N) 
\\
 \langle A \rangle &=& \frac{ 5 \pi}{192}\,   \sqrt{\frac{3}{2}}\,  N^3 +O(N^{5/2}).
\eea
In Figs. (\ref{figp}) and (\ref{figa}), we present our data respectively for the 
average perimeter $\langle L 
\rangle$ and the average area $\langle A \rangle$ of the convex hull, as a 
function of $N$.
Both the scaling and the numerical prefactor of the leading behavior in $N$ are 
recovered. The agreement between the simulation and the analytical results is 
excellent.

Furthermore, something rather interesting can be remarked on Fig. \ref{figV}. 
For 
large $N$, the leading behavior of the average number of vertices seems
to be $\langle 
V\rangle\sim \sqrt{N}$. Numerically we find
\be
\langle V\rangle \sim c_1 \sqrt{N} + c_2 +O(1/\sqrt(N)).
\ee
A best fit gives $c_1=1.735 \pm 0.01$ and $c_2=2.6 \pm 1.0$.  
We were not able to derive this result analytically and it remains
an interesting open question.

\section{Summary and Conclusion}

In summary, in this paper we have computed analytically the mean perimeter $\langle L(T)\rangle$
and the mean area $\langle A(T)\rangle $ of the convex hull of a two dimensional random acceleration 
process of duration $T$, starting at the origin with zero initial velocity. 
We have used an exact mapping, via Cauchy's formulae, that relates the computation of
the perimeter and the area of the convex hull of a two dimensional stochastic process to
computing the extremal statistics associated with the corresponding one dimensional
component stochastic process. Physically our results describe the average shape of
a semi-flexible ideal polymer chain in two dimensions.

We have also presented numerical results that suggest that the mean number of vertices $V$
of the convex hull associated to the discrete-time version of the random acceleration process 
up to $N$ steps scales as $\langle V(N)\rangle \approx c_1 \sqrt{N} + c_2 + O(1/\sqrt{N})$ for large $N$.
Proving this result and the computation of the constants $c_1$ and $c_2$ remain 
a challenging open problem.

It is interesting to compare our exact results for the convex hull associated
with a two-dimensional random acceleration process with those for the convex
hull assocaited to a planar Brownian motion. In particular one notes 
that the ratio $r= \langle A\rangle/{\langle L\rangle }^2$ has 
very different values for the two processes. 
For a random acceleration process, using our exact results in
Eqs. (\ref{ramper1}) and (\ref{ramarea1}), we get
$r=(5/192)\sqrt{2/3}=0.02126\dots$. The corresponding ratio
for a free
Brownian motion, using results mentioned in the introduction, 
is $r=1/16=0.0625$ which is larger than the random acceleration process.
The geometrical implication of this result is that the convex hull
associated with the
Brownian motion is, roughly speaking, more
space filling than the one associated with the random acceleration
process. In the latter case, the convex hull tends to be more
`elongated' as evident from a typical representation in Fig. 
(\ref{typical.fig}). This is somewhat expected from
the statistical mechanical consideration of the semi-flexible polymer 
chains described by the random acceleration process: the typical 
configuration of such a curvature-driven chain tends to be 
elongated 
as it costs more energy to bend or curve them as compared to a typical 
Brownian trajectory of a Rouse chain.    

There are several directions in which our work can be possibly extended. 
In this paper, we managed to compute only the mean perimeter
and the mean area of the convex hull of the random acceleration process. It remains a challenging 
problem to compute the fluctuations or even
the full distribution of these observables associated with the convex hull of the 
random acceleration process.
The mapping we discussed in this paper
is quite general and holds for arbitrary two dimensional stochastic process. It would be interesting
to exploit this mapping to compute the mean perimeter and the mean area of the convex hull
of other interesting two dimensional 
non-Markovian stochastic processes. For instance, for the process $d^n\vec r/dt^n= \vec \eta(t)$
(where $n=1$ and $n=2$ correspond respectively to the Brownian motion and the random acceleration process)
with $n>2$, it would be interesting to see if one can compute the statistics of the associated
convex hulls. 

\appendix

\section{The moments of the maximum value distribution} \label{moments}

Consider a one dimensional random acceleration process $x(t)$ in the time interval $[0,T]$. At 
time $t=0$ the process starts from $x_0=0$ with zero velocity, $v_0=0$. Let $m(T)$ 
denote the global maximum value achieved by $x(t)$ during the interval $[0,T]$.
We wish to compute the cumulative probability distribution of the maximum, i.e.,
${\rm Prob}\left[m(T)<z\right]$. Clearly, this is the probability that the 
value of the process, starting initially at $x=0$, does not exceed the level $z$.
In other words, ${\rm Prob}\left[m(T)<z\right]= {\rm Prob}\left[ x(t)<z,\,\, {\rm for\,\, all\,\, 
0\le t\le T}\right]$.   
We then introduce the change of variable $\tilde{x}(t)=z -x(t)$.
The process $\tilde{x}(t)$ is also a random acceleration 
process, starting from $z$ (with zero velocity). The probability that $m(T)$ is 
smaller than $z$ can then be expressed in terms of the survival probability of the new 
process $\tilde{x}(t)$, i.e., the probability $q(z,0,T)$ that the process $\tilde{x}(t)$ stays
positive over the initerval $t\in [0,T]$ given that it started from the initial value
$\tilde{x}(0)=z$. We then write
\begin{equation} 
\mbox{Prob}\left[ m(T) < z \right] = \mbox{Prob}\left[ \{ x(t) \} 
< z \right] = \mbox{Prob}\left[ \{ \tilde x(t) \} >0 \right] =q(z,0,T). 
\end{equation} 
Note that $q(0,0,T)=0$ and $q(z\to\infty, 0,T)=1$.

The moments of the maximum $m$ can be expressed in terms of the 
survival probability:
\begin{equation}
\label{n-thmom}
\langle m^n(T) \rangle =\int_0^\infty d z \, z^n \frac{d q}{d z}= 
-\int_0^\infty d z\,  z^n \frac{d }{d z} \left[1-q\right]= 
n \int_0^\infty d z\,  z^{n-1} \left[ 1-q(z,0,T)\right]
\end{equation}
where we have used integration by parts and the boundary condition $q(z\to \infty, 0, T)=1$. 
Let $\tilde{q}(z,0,s)$ denote the Laplace transform of the survival porbability:
\begin{equation}
 \tilde{q}(z,0,s)=\int_0^\infty  d T  \, e^{- s T} \, q(z,0,T).
\end{equation}
This quantity can be expressed as an integral~\cite{Burkhardt,Burkhardt5}:
\begin{equation}\label{q}
\frac{1}{s} - \tilde{q}(z,0,s) =\int_0^\infty \frac{dF}{F^{5/3}} 
\, e^{-F z} \mbox{Ai}\left(\frac{s}{F^{2/3}}\right) 
\left[  1+\frac{\Gamma\left(-\frac{1}{2}, \frac{2}{3} \frac{s^{3/2}}{F} 
\right)}{4 \sqrt{\pi}}  \right],
\end{equation}
where $ \Gamma\left(-\frac{1}{2}, y \right)$ is the 
incomplete Gamma function. We can thus introduce the Laplace transform of 
the moments in Eq. (\ref{n-thmom}) and obtain 
\begin{equation*}
\nonumber
\int_0^\infty  d T \, e^{- s T} \, \langle m^n(T) \rangle 
= n \int_0^\infty d z\,  z^{n-1} \int_0^\infty \frac{dF}{F^{5/3}} \, 
e^{-F z} \mbox{Ai}\left(\frac{s}{F^{2/3}}\right) 
\left[  1+\frac{\Gamma\left(-\frac{1}{2}, \frac{2}{3} 
\frac{s^{3/2}}{F} \right)}{4 \sqrt{\pi}}  \right].
\end{equation*}
Performing the integral over $z$ we get
\begin{equation*}
\int_0^\infty  d T  \, e^{- s T} \, \langle m^n(T) \rangle =  
\Gamma(n+1) \int_0^\infty \frac{dF}{F^{5/3 +n}} \,  
\mbox{Ai}\left(\frac{s}{F^{2/3}}\right) 
\left[  1+\frac{\Gamma\left(-\frac{1}{2}, \frac{2}{3} \frac{s^{3/2}}{F} 
\right)}{4 \sqrt{\pi}}  \right].
\end{equation*}

Next we introduce a change of variable  $y=2 s^{3/2}/(3 F)$,
\begin{equation*}
\int_0^\infty  d T \, e^{- s T} \, \langle m^n(T) \rangle =  
\frac{\Gamma(n+1)}{s^{\frac{3}{2} n +1}}  
\left(\frac{3}{2}\right)^{n +\frac{2}{3}}  
\int_0^\infty  d y y^{n-\frac{1}{3}} \,  
\mbox{Ai}\left[   \left(\frac{3 y}{2}\right)^{2/3}\right] 
\left[  1+\frac{\Gamma\left(-\frac{1}{2}, y \right)}{4 \sqrt{\pi}}  \right]. 
\end{equation*}
As a result, the $s$ dependence of the r.h.s is very simple, just proportional
to $s^{-3n/2-1}$ and hence its inverse Laplace transform (with respect to $s$) can
be done trivially to give
\begin{equation}
 \langle m^n(T) \rangle = \left\{ \left(\frac{3}{2}\right)^{n -\frac{1}{3}}
 \frac{\Gamma(n)}{\Gamma(\frac{3}{2} n)} \int_0^\infty  d y \, 
y^{n-\frac{1}{3}} \,  \mbox{Ai}\left[   \left(\frac{3 y}{2}\right)^{2/3}\right] 
\left[  1+\frac{\Gamma\left(-\frac{1}{2}, y \right)}{4 \sqrt{\pi}}  \right] 
 \right\}  T^{\frac{3}{2} n}. 
\end{equation}

At this point, it turns out to be useful to express the Airy function in 
terms of the modified Bessel
function $K_\nu(z)$ using the following identity \cite{GR}:
\begin{equation}
  \mbox{Ai}\left[   \left(\frac{3 y}{2}\right)^{2/3}\right] =  
\frac{1}{\sqrt{3} \pi} \left(  \frac{3}{2}  \right)^{\frac{1}{3}}\, y^{1/3}\, 
\mbox{K}_{\frac{1}{3}}(y).
\end{equation}
Then we have
\begin{equation}
\label{m1}
 \langle m^n(T) \rangle =  \left(\frac{3}{2}\right)^{n} 
\frac{\Gamma(n)}{ \sqrt{3} \pi \Gamma(\frac{3}{2} n)} \,  
I^{(n)}\,  T^{\frac{3}{2} n}
\end{equation}
where
\begin{equation}
  I^{(n)}  =  \int_0^\infty  d y \, y^n \, 
\mbox{K}_{\frac{1}{3}}(y) \left[  1+\frac{\Gamma\left(-\frac{1}{2}, 
y \right)}{4 \sqrt{\pi}} \right].
\end{equation}

We now show that the integral $I^{(n)}$ can be computed for any non-negative
integer $n$.
Starting from the definition of the incomplete $\Gamma$ function, 
$\Gamma\left(\alpha, y \right) = \int_y^\infty d t \, t^{\alpha+1} \, e^{-t}$, we 
first use the following decomposition formula valid for any non-negative integer $n$,
\begin{equation}
\frac{\Gamma\left(-\frac{1}{2}, y \right)}{4 \sqrt{\pi}} = 
\frac{1}{2 \pi} \left[  (-1)^n \Gamma(n+\frac{3}{2})   
\Gamma(-n-\frac{1}{2},y)  -    e^{-y} \sum_{k=1}^n (-1)^k  
\frac{\Gamma(k+\frac{1}{2})}{y^{k+\frac{1}{2}}}    \right],
\end{equation}
which can be easily obtained by repeated integration by parts.
We then split the main integral $I^{(n)}$ into three parts 
$I^{(n)} =I_1+I_2+I_3$ with
\bea
I_1=\int_0^\infty \, dy \,  y^{n} \mbox{K}_{\frac{1}{3}}(y) \nonumber \\
I_2=(-1)^n  \Gamma(n+\frac{3}{2})     
\int_0^\infty  \frac{d y}{2 \pi}  \, y^n \, 
\mbox{K}_{\frac{1}{3}}(y)   \Gamma(-n-\frac{1}{2},y) \nonumber \\
I_3= -  \sum_{k=1}^n (-1)^k \Gamma(k+\frac{1}{2})  
\int_0^\infty  \frac{d y}{2 \pi}  \, y^{n-k-\frac{1}{2}}\, e^{-y} 
\, \mbox{K}_{\frac{1}{3}}(y) 
\eea

\subsubsection{Evaluation of $I_2$}
 
Using the  identity \cite{GR},
\begin{equation*}
  \Gamma(n+\frac{3}{2})  \Gamma(-n-\frac{1}{2},y)  =    
\frac{e^{-y}}{ y^{n+1/ 2}} \int_0^\infty \, dt \, \frac{e^{- t} t^{n+1/2}}{t +y},
\end{equation*}
and the formula \cite{GR},
\begin{equation*}
 \int_0^\infty \frac{d y}{\sqrt{y}} \frac{e^{-y} 
\mbox{K}_\nu(y)}{y+t}=\frac{\pi \, e^t  \, 
\mbox{K}_{\frac{1}{3}}(t)}{\sqrt{t} \cos(\nu \pi)},
\end{equation*}
 $I_2$ can be recast:
\begin{equation*}
I_2= (-1)^n\int_0^\infty \, dt \, e^{- t} t^{n+1/2}\int_0^\infty   
\frac{dy}{2 \pi \sqrt{y}} \, \frac{e^{-y}}{t+y} \, \mbox{K}_{\frac{1}{3}}(y)
\end{equation*}
and the integral over $y$ can be performed
\begin{equation*}
I_2= (-1)^n \int_0^\infty \, dt \,  t^{n} \mbox{K}_{\frac{1}{3}}(t)=  (-1)^n 
I_1
\end{equation*}
\subsection{Evaluation of $I_1$}
The integral $I_1$ can be computed \cite{GR}
\begin{equation*}
I_1= \sqrt{3} \pi \left(\frac{2}{3}\right)^n 
\frac{\Gamma(\frac{3}{2}n)}{\sqrt{3^n}\Gamma(\frac{1}{2}n)}
\end{equation*}
\subsubsection{Evaluation of $I_3$}
Using the integral \cite{GR}
\begin{equation*}
\int_0^\infty   d y  \, y^{\mu-1}   \, e^{-y} \, 
\mbox{K}_{\frac{1}{3}}(y)=\frac{\sqrt{\pi}}{2^\mu} \frac{\Gamma(\mu+1/3)  
\Gamma(\mu-1/3)   }{\Gamma(\mu+1/2)  }
\end{equation*}
with $\mu=n-k+1/2$ and the change of variable $j=n-k$ we get
\begin{equation*}
I_3=   \frac{(-1)^{n-1}}{\sqrt{8 \pi}} S_n 
\end{equation*}
where
\begin{equation}
S_n= \sum_{j=0}^{n-1}  \frac{\Gamma(j+5/6)  \Gamma(j+1/6)  
\Gamma(n-j+\frac{1}{2}) }{ (-2)^j  \Gamma(j+1)  }.
\label{SN}
\end{equation}
This sum can be performed using Mathematica but it gives a rather cumbersome expression.
Hence we prefer to keep this as a discrete sum in Eq. (\ref{SN}).

\subsection{The final result}

Putting together all these integrals in Eq. (\ref{m1}) we get
\begin{equation}
\label{m2}
 \langle m^n(T) \rangle = \left[\frac{(-1)^n+1}{3^{\frac{n}{2}}} 
\frac{\Gamma(n)}{\Gamma(\frac{n}{2})}-\frac{1}{\sqrt{24 \pi^3}} 
\left(-\frac{3}{2}\right)^n  \frac{\Gamma(n)}{\Gamma(\frac{3}{2} n)} 
S_n \right] T^{\frac{3 n}{2}}
\end{equation}
where $S_n$ is given in Eq. (\ref{SN}).
For example, for the first four moments are given explicitly by
\bea
\langle m(T) \rangle = \sqrt{\frac{3}{8 \pi}} \, T^{\frac{3}{2}} \label{mean}\\
\langle m^2(T) \rangle = \left( \frac{2}{3}-\frac{103}{384}
\sqrt{\frac{3}{2}}\right)\, T^{3}  \label{second}\\
\langle m^3(T) \rangle =\frac{1091}{864} \,  \sqrt{\frac{3}{8 \pi}}   
\, T^{\frac{5}{2}}  \label{third} \\
\langle m^4(T) \rangle =\left( \frac{4}{3}-
\frac{5769947}{10616832}\sqrt{\frac{3}{2}}\right)\, \, T^{5}.  \label{four}
\eea

\end{document}